\documentclass[twocolumn,pre, showpacs]{revtex4}
\usepackage{amsmath}
\usepackage{graphicx}
\usepackage{multirow}
\usepackage{color}

\begin{document}
\title{A First Passage Time Analysis of Atomic-Resolution Simulations of the Ionic Transport in a Bacterial Porin}
\author{Carles Calero $^1$}
\email{ccalero@icmab.es}
\author{Jordi Faraudo $^1$}
\author{Marcel Aguilella-Arzo $^2$}

\affiliation{$^1$ Institut de Ci\`encia dels Materials de Barcelona (ICMAB-CSIC), Campus de la UAB, E-08193 Bellaterra, Spain.
	     $^2$ Biophysics Group, Department of Physics, Universitat Jaume I, 12080 Castell\' o, Spain.}

\begin{abstract}
We have studied the dynamics of chloride and potassium ions in the interior of the OmpF porin under the influence of an external electric field. From the results of extensive all-atom molecular dynamics simulations of the system we computed several first passage time (FPT) quantities to characterize the dynamics of the ions in the interior of the channel. Such FPT quantities obtained from MD simulations demonstrate that it is not possible to describe the dynamics of chloride and potassium ions inside the whole channel with a single constant diffusion coefficient. However, we showed that a valid, statistically rigorous, description in terms of a constant diffusion coefficient $D$ and an effective deterministic force $F_\text{eff}$ can be obtained after appropriate subdivison of the channel in different regions suggested by the X-ray structure. These results have important implications for popular simplified descriptions of channels based on the 1D Poisson-Nernst-Planck (PNP) equations. Also, the effect of entropic barriers on the diffusion of the ions is identified and briefly discussed. 
\end{abstract}

\maketitle

\section{Introduction}
\label{sec:Introduction}

Understanding transport of charged solutes across the cell membrane is a problem of paramount importance in biophysics, since it is crucial to regulate many cell functions \cite{Hille}. This task is undertaken by transmembrane channels, so the characterization of the permeation of charged particles, ions in particular, is essential. In addition, recently there has been a renewed interest to comprehend transport through biological nanochannels for possible applications in biotechnology \cite{Maglia}. 

Bacterial porins are macromolecular proteins located at the outer membrane that enable the diffusion of small molecules through the lipid bilayer. As porins are well characterized, both structurally and functionally, they represent model systems to study transport through biological nanochannels. In particular, the ionic transport properties of the OmpF porin have been extensively studied since the determination of its X-ray structure \cite{Cowan}. 

The simplest description of the complex problem of ionic transport across channels is based on the classical diffusion-advection transport equations:
\begin{equation}
\label{cons}
\frac{\partial c(z,t)}{\partial t}+\frac{\partial J(z,t)}{\partial z}=0 
\end{equation}
\begin{equation}\label{NP}
 J(z,t) = -D\left(\frac{dc(z,t)}{\partial z} - \frac{F(z)}{k_BT}c(z,t)  \right)\,,
\end{equation}
Here, $J(z,t)$ is the flux of particles of a given ion of charge $q$ and $c(z,t)$ its concentration profile, $D$ is a constant diffusion coefficient characterizing the brownian motion (with a value in principle different from bulk diffusion coefficient) and $F(z)$ is a deterministic force. In the classical Nernst-Planck equations, the deterministic force arises from the mean-field electrostatic potential, $F(z)=-qd\phi(z)/dz$ \cite{Hille}, whereas in more advanced approaches it may contain steric or entropic forces \cite{Jacobs, Zwanzig, Reguera}.

In some cases, integrated solutions of Eqs.\,(\ref{cons},\ref{NP}) under suitable boundary conditions and appropriate assumptions are employed. The Goldman-Hodkin-Katz equation or the Henderson equation are well-known examples, being extensively used in the literature to understand the conductance, reversal potential, and other measurable quantities of nanochannels \cite{Hille}. In other cases, it is solved self-consistently in combination with Poisson's equation of electrostatics, forming the so-called Poisson-Nernst-Planck (PNP) equations. This approach allows the analysis of electrically charged nanochannels in a consistent fashion \cite{Chung,Chen1997, Chen1999, Tang, Roux1}. In all cases, a continuous mean-field approximation is assumed, i.e. ions are treated not as discrete entities but as continuous charged densities that represent the space-time average of the microscopic motion of the ions. The diffusion coefficient of ions, $D$, characterize their motion within the channel, representing a spatial average over the region considered. In analyzing experiments, diffusion coefficients are often given bulk values when the comparison with experiment is qualitative or are treated as free parameters to adjust by fitting experimental data \cite{Lidon, Chen1997, Chen1999, Tang}. \\

In recent years, the increase in computer power and the development of new algorithms have permitted the study of ion transport through certain nanochannels employing all-atom Molecular Dynamics (MD) simulations \cite{Aksimentiev, Ulrich2009, Faraudo}. Such analysis captures all the atomic detail of the system (including the water molecules present), which in some cases is necessary to properly account for the interactions of the ions with the interior of the channel. However, these MD simulations still remain computationally very expensive, so its use is hindered by limitations in the size of the system ($\sim 10-100$ nm) and the duration of the processes ($\sim 100$ ns) that such calculations are capable of tackling. Nevertheless, even for the analysis of larger systems or longer processes, all-atom MD simulations can be used to obtain parameters that coarser (less detailed) models employ in their description of the system. In particular, MD simulations can be used to obtain diffusion coefficients of ions in the interior of nanochannels, which are central input parameters in many theoretical descriptions of ion permeation (like, e.g., Brownian Dynamics simulations \cite{Roux2001, Roux2002b, Schirmer, Phale} or approaches based on the Nernst-Planck equation \cite{Bolintineanu}). Tieleman and Berendsen \cite{Tieleman} characterized the diffusive dynamics of water molecules in the OmpF porin by using the Einstein relation for the mean square displacement (MSD). They calculated local diffusion coefficients which they averaged out over slices perpendicular to the axis of the pore to obtain the variation of the diffusion coefficient along the axis of the channel. A similar approach was used with the OmpF porin \cite{Roux2002} and other channels\cite{Aksimentiev, Allen} to determine the dependence of ionic diffusion coefficients along their axial direction. Such approaches, however, do not provide global diffusion coefficients to describe the entire channel, and it is not clear how they can be determined. To circumvent that limitation, Hijkoop et al. \cite{Hijkoop}, inspired on the work of Munakata and Kaneko \cite{Munakata}, used a method based on a first passage time (FPT) analysis to compute a global diffusion coefficient of water in the interior of an artificial (uncharged) OmpF porin. Their analysis, however, ignores the influence of free energy barriers and wells on the local movements of water molecules at a specific position along the axis of the pore.

In this paper, we use a method based on a first passage time (FPT) analysis of the particles' trajectories to study the transport of potassium and chloride ions through the OmpF porin. This procedure allows us to obtain the effective diffusion coefficients of the ions in the channel from all-atom MD simulations \cite{VanKampen, Redner, Hijkoop}. On the one hand, from the trajectories of the ions provided by the simulations one can calculate several FPT quantities that characterize their passage through the channel. Note that such quantities are produced taking into account all the complexity of the all-atom description. On the other hand, it is possible to derive analytical expressions of such FPT quantities for a model based on the 1D diffusion-advection equations with constant parameters, Eqs.\,(\ref{cons},\ref{NP}). The comparison of the FPT quantities obtained from the MD simulations with the analytical formulas give answer to two relevant questions for the characterization of the channel. First, it determines whether the one-dimensional description of the dynamics of the ions in the channel based on Eqs.\,(\ref{cons},\ref{NP}) is valid. Second, in case such a description is accurate, we can obtain the diffusion coefficients and effective forces that best define the channel by fitting the FPT quantities calculated from the MD simulations with their corresponding analytical expressions. 

The rest of the paper is organized as follows. In section \ref{sec:OmpF + simulation methods} we describe the structure of the OmpF porin and discuss the model and the details of the MD simulations. In section \ref{sec:First Passage Approach}, we define the first passage quantities which will be considered and give the corresponding analytical expressions for the 1D model based on Eq.\,(\ref{cons},\ref{NP}). In section \ref{sec:Results} we provide the results of the analysis. We discuss the FPT quantities obtained for the whole channel and also study the FPT properties of different regions within the channel. From this analysis, effective diffusion coefficients and forces are obtained for the different regions. In section \ref{sec:Conclusions}, the conclusions of the study are given, as well as a table with the obtained diffusion coefficients that characterize the ionic transport of the channel.

\section{System and simulation methods}
\label{sec:OmpF + simulation methods}

\begin{figure}[ht!]
   \begin{center}
      \includegraphics[width=7.5cm]{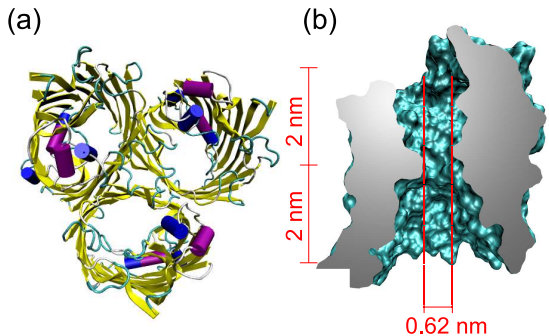}
      \caption{(Color online) Protein structure derived from its X-ray structure \cite{Cowan}. (a) Top view of the OmpF porin. (b) Cross section of a single monomer, showing its dimensions. Figure produced using VMD \cite{VMD}}
      \label{channel}
   \end{center}
 \end{figure}

The biological nanochannel considered here, the outer membrane protein F (OmpF porin), is a transmembrane channel located at the outer membrane of the bacteria {\it Escherichia coli}. The structure of the WT OmpF channel is illustrated in Figure \ref{channel}. The crystallographical structure of the wild type OmpF channel is well known from long ago \cite{Cowan,PDB}. It is made of three identical monomerical nanopores, each one assembling into a large 16-stranded antiparallel $\beta$-barrel structure enclosing the transmembrane pore (Figure \ref{channel}). Each aqueous pore has a diameter between 1-4 nm, being constricted around half-way through the membrane by a long loop. It is a relatively wide channel in the sense that it allows the simultaneous permeation of both cations and anions (in {\it hydrated} form) and also the passive diffusion of larger dipolar molecules. The transport of monovalent ions and molecules across this channel is known to be dominated by electrostatics, as demonstrated by studies considering mutations of the channel (with different pore sizes and different electrostatic charges) and different permeating specimens of different sizes and charge distributions  \cite{Phale,Rostovtseva2002,Danelon2006}. 

\begin{figure}[ht!]
   \begin{center}
      \includegraphics[width=8.5cm]{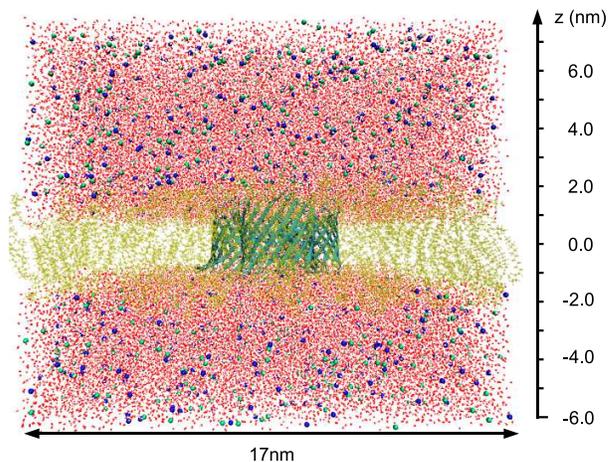}
      \caption{(Color online) Snapshot from the MD simulation, in which the dimensions of the simulation box have been added. In the picture, two slabs of 1M KCl aqueous solution are separated by a POPC lipid bilayer in which the trimer OmpF porin is inserted. To favor the passage of ions, an electric field along the negative $z$ axis is applied. For clarity, only a fraction of the lipids of the membrane are represented. Figure produced using VMD \cite{VMD}.}
      \label{snapshot}
   \end{center}
 \end{figure}

In this paper we study the transport of potassium and chloride ions through the OmpF porin using a FPT approach from results of all-atom molecular dynamics simulations previously published in Refs. \cite{Faraudo, Marcel}. For the sake of completeness, we summarize here the most important features of the simulation. The system is composed of the OmpF porin inserted in a POPC membrane in contact with a 1M KCl aqueous ionic solution, see Fig.\ref{snapshot}. The whole system is thermalized at $T_r = 296$K. The coordinates of the OmpF porin are obtained from its X-ray structure \cite{Cowan}. The protonation states employed for the titatrable residues of the protein are the same as those specified in Ref. \cite{Varma}, which results in an overall charge of $-11$e per monomer. The molecular dynamics simulations were carried out with NAMD version 2.6 \cite{NAMD}. We used the force-field for protein-lipid simulations supplied by NAMD2, which is a combination of the CHARMM22 force field, parameterized to describe protein systems, and the CHARMM27 force field\cite{Charmm}, parameterized to describe lipid systems in aqueous media. Water was also modeled after the CHARMM force field \cite{Charmm}, which describes water using a slightly modified version of the standard TIP3P model. Within this force field, ions are described as charged Lennard-Jones spheres. All simulation details can be found in Refs.\cite{Faraudo, Marcel}.

To study the ionic transport properties of the channel, the whole system is under the action of a uniform electric field perpendicular to the lipid bilayer (along the negative $z$ axis, see Fig.\ref{snapshot}). The magnitude of the field is $E_{ext} = 14.22$mV/nm, which creates a potential drop of $\sim 200$ mV across the membrane+channel system. From a computational perspective, the high electrolyte concentration, of approximately 1M, is convenient, since the large number of ions included in the simulation significantly helps to obtain statistically meaningful averages of the ion properties inside the nanochannel. Such high concentrations, however, are not unrealistic. Similar concentrations are often used in experiments to measure ion fluxes and selectivity properties of ion channels \cite{Lidon}. The total simulated time is $\sim 30$ns, considerably longer than previous MD simulations done of the OmpF porin \cite{Roux2002, Ulrich2009}, which also helps to improve the statistical averages of ion properties inside the channel. 

\section{First Passage Time (FPT) Approach}
\label{sec:First Passage Approach}

Let us consider a diffusing Brownian particle under the action of an external force in a one dimensional interval $[0,L]$. To characterize its dynamics in the interval one can compute first passage quantities, which deal with the event of the particle reaching for the first time any of the boundaries (first passage event). The central magnitude that characterizes these events (from which other first passage quantities can be calculated) is the distribution of first passage times, given a starting position of the particle within the interval. This is obtained from the probability that the particle reaches for the first time any of the boundaries at a given time $t$. This first passage problem is, thus, approximately equivalent to the problem of a diffusing particle in an interval $[0,L]$ with absorbing boundaries \cite{VanKampen, Redner}. 

\subsection{Calculation of FPT quantities from MD results}

The analysis based on a first passage approach is well suited to characterize the dynamics of the ions in the channel from the results of molecular dynamics simulations. One can easily calculate the different first-passage quantities characterizing the ions dynamics inside a given interval within the channel from inspection of the individual trajectories provided by the simulations' results.

A first passage quantity that we employ in our analysis is the {\it Hitting Probability}. It is defined as the probability that a particle which starts at an initial position $z_0$ reaches for the first time either the lower boundary at $z=0$, $P_-(z_0)$, or the upper one at $z=L$, $P_+(z_0) = 1-P_-(z_0)$. The dependence of $P_{\pm}(z_0)$ on $z_0$ is determined by the presence of free energy barriers and external forces acting on the ions. The calculation of the hitting probability from the MD trajectories of the ions is simple. At each timestep of the simulation we identify the ions contained within the layer $[z_0 - \lambda/2, z_0 +\lambda/2]$, with $\lambda = 0.1$ nm (thiner layers were tested with identical results). We then trace all those particles until they reach for the first time any of the virtual boundaries of the interval, moment at which they loose their identity (although they are not physically removed from the system). By keeping track of the number of ions that reach each boundary we can compute, once we have gone through all the simulation, the hitting probabilities of each type of ion for a given initial position $z_0$. The same procedure is repeated for different starting points $z_0$ (separated by $\lambda$) until the whole interval is covered to obtain the hitting probabilities from MD simulations, $P_{\pm}^{(MD)}(z_0)$.

Another magnitude of interest in the application of the FPT approach to the analysis of ionic dynamics in the channel is the {\it Mean Exit Time} (MET), $T(z_0)$. It is defined as the average time taken for a particle to reach for the first time any of the boundaries given that its initial position is $z_0$. To calculate the MET from the MD simulations results we divide the system into layers of width $\lambda$ as done before, and at each timestep identify the ions within the layer centered at $z_0$. Those particles are tracked until they reach either boundary and the time elapsed is recorded. From the collection of residence times of the particles starting at $z_0$ we compute the average time to reach either boundary. This is repeated for all $z_0$ in the interval to obtain the MET from MD simulations, $T^{(MD)}(z_0)$.

The other FPT quantity that we use in the characterization of ion dynamics in the channel is the {\it Survival Probability}, $S_p(t)$. It gives the probability at time $t$ of finding a particle within the domain $[0, L]$ that will cross the domain from one boundary to the other without crossing the entrance at intermediate times. For the calculation of this quantity, at each timestep of the simulation we identify those ions that enter the considered interval and, for each particle, we record  the time of entrance into the interval and a tag which specifies the boundary of entrance. As done in the previous cases, we track those particles until they reach any of the boundaries, where they loose their identity. If the boundary of exit is the same as the boundary of entrance the trajectory is discarded. If the particle exits through the opposite boundary, the exit time is recorded so we can compute the time taken by that particle to cross the interval. From the collection of the crossing times it is then straightforward to calculate the survival probability from MD simulations, $S_p^{(MD)}(t)$. In the presence of an external effective force dragging the ions, there are many more events of ions crossing the interval in the direction of the dragging force than in the opposite direction. In order to provide statistically meaningful results, only the survival probability corresponding to the passage of ions in the direction of the dragging force will be considered.

\subsection{Analytical results for a simplified model}

The FPT quantities obtained from the MD simulations can be used to check the validity and to extract the parameters of a certain theoretical model. Based on such a model, one can calculate the corresponding theoretical FPT quantities $P_{\pm}^{(th)}(z_0; \{\alpha_n\}), T^{(th)}(z_0; \{\alpha_n\}), S_p^{(th)}(t; \{\alpha_n\})$, where $\{\alpha_n\}$ are the parameters of the model. By comparing such theoretical quantities with those computed from MD simulations, $P_{\pm}^{(MD)}(z_0), T^{(MD)}(z_0), S_p^{(MD)}(t)$, we can test the applicability of the model and also, under favorable conditions, determine the parameters $\{\alpha_n\}$ that define such model. Note that the parameters thus obtained contain in an effective fashion all the information on the geometry and interactions of the all-atomic simulation.

Here, we consider the simplest diffusion model based on ions characterized by a constant diffusion coefficient $D$ moving under the influence of a constant effective force $F_\text{eff}$. For such a simple model, the classical transport Eqs. \,(\ref{cons},\ref{NP}) apply and simple analytical expressions for the FPT quantities can be deduced in terms of the parameters $D$ and $F_\text{eff}$. These analytical expressions are derived by noting that the one-dimensional FPT problem is approximately equivalent to the problem of a diffusing particle in a 1D interval with absorbing boundaries. For detailed derivations, the reader is refereed to standard textbooks such as Ref. \cite{Redner}. Here we limit ourselves to list the results relevant for our analysis. 

The Hitting Probabilities are given, in terms of $D$ and $F_\text{eff}$, by \cite{Redner}
\begin{eqnarray}\label{hitprob}
P_-^{(th)}(z_0) & = & \frac{e^{-F_\text{eff}\,z_0/k_BT} - e^{-F_\text{eff}L/k_BT}}{1- e^{-F_\text{eff}L/k_BT}} \nonumber \\
P_+^{(th)}(z_0) & = & \frac{1 - e^{-F_\text{eff}\, z_0/k_BT}}{1- e^{-F_\text{eff}L/k_BT}}\,. 
\end{eqnarray}
Note that $P_{\pm}^{(th)}(z_0)$ depend solely on one of the parameters of the model, $F_\text{eff}$. Consequently, the comparison of Eqs.\,(\ref{hitprob}) with the results from the MD simulations, $P_{\pm}^{(MD)}(z_0)$, is well suited to obtain the effective force acting on the ions.

Within the considered theoretical model, the Mean Exit Time reads \cite{Redner}
\begin{equation}\label{MET}
 T^{(th)}(z_0) = \frac{L}{D} \frac{k_BT}{F_\text{eff}} \frac{1 - e^{\frac{-F_\text{eff}\, z_0}{k_BT}}- \frac{z_0}{L}[1- e^{\frac{-F_\text{eff}L}{k_BT}}]} {1- e^{\frac{-F_\text{eff}L}{k_BT}}}\,.
\end{equation}

Finally, the analytical expression for the survival probability, $S_p^{th}(t)$, can also be deduced from the diffusion-advection equations, Eqs.\,(\ref{cons},\ref{NP}), \cite{Redner}:
\begin{equation}\label{SurvP}
 S_p^{(th)}(t) \propto \sum_{n =1}^{\infty}  \frac{(-1)^{n-1} n^2}{n^2 + (F_\text{eff}L/2\pi k_BT)^2} e^{-\alpha_n t} \,,
\end{equation}



with $\alpha_n = ({D}/{L^2})\left[n^2\pi^2 + (F_\text{eff}L/2k_BT)^2 \right]$. At times $t \gg 1/\alpha_2$, $S_p^{(th)}(t)$ is dominated by the first term, with $\alpha_1 = ({D}/{L^2})\left[\pi^2 + (F_\text{eff}L/2k_BT)^2 \right]$, which directly depends on the diffusion coefficient.

In the analyses done below, based on the comparison of the FPT results from the MD simulations with Eqs.\,(\ref{hitprob}), (\ref{SurvP}) and (\ref{MET}),  the interval length $L$ needs to be considered slightly longer than its actual length in the MD simulations \cite{Hijkoop}. This extra distance, which appears to be dependent on the type of particle and the interval, has been studied and identified before as the {\it Milne extrapolation length} \cite{Hijkoop, Marshall}. The necessity for such a (very small) correction emerges because the absorbing boundary conditions assumed in \cite{Redner} to derive Eqs.\,(\ref{hitprob}),  (\ref{MET}) and (\ref{SurvP}) are not exact for the virtual interval considered in the analysis of the simulations.

\section{Results and Discussion}
\label{sec:Results}

\subsection{Test case: Bulk Electrolyte}
To check the validity of the method described above, we analyzed first the dynamics of the ions in bulk, far from the channel and membrane, which we know is describable using a 1D Nernst-Planck equation (Eqs.(\ref{cons},\ref{NP})) with constant parameters and whose results can also be obtained from the analysis of the mean square displacement (MSD).\\

For this analysis we have selected a 2nm-thick slab of electrolyte far enough from the channel and membrane, $-6.0$nm$ < z < -4.0$ nm (see Fig.\,\ref{snapshot}). The delimiting planes of the slab perpendicular to the $z$-axis act as virtual absorbing boundaries, while in the transverse directions ($x$, $y$) periodic boundary conditions are applied.

The protocol that was followed to obtain the parameters of the theoretical model, $D$ and $F_\text{eff}$, is as follows. The effective force parameter $F_\text{eff}$ is determined first from fitting the hitting probabilities of reaching the upper boundary of the specified region, $P_+^{(MD)}(z_0)$, with Eqs.\,(\ref{hitprob}). The dependence of these probabilities on the initial position of the particle is determined only by the effective force $F_\text{eff}$, being a straight line for $F_\text{eff} = 0$ and increasing its curvature as the parameter grows (concave for $F_\text{eff} >0$ and convex for $F_\text{eff} <0$). Second, the diffusion coefficients are obtained from the comparison of the MD results for the survival probability, $S_p^{(MD)}(t)$, with Eq.(\ref{SurvP}) at $t \gg 1/\alpha_2$. The slope of $\log[S_p^{(MD)}(t)]$ directly gives the diffusion coefficient, provided that $F_\text{eff}$ is known. An alternative way to obtain the diffusion coefficients that is used to check the consistency of the results consists of the comparison of the computed mean exit times, $T^{(MD)}(z_0)$, with the analytical expression Eq.\,(\ref{MET}). 

\begin{figure}[ht!]
   \begin{center}
        \includegraphics[angle=0,width=8.5cm]{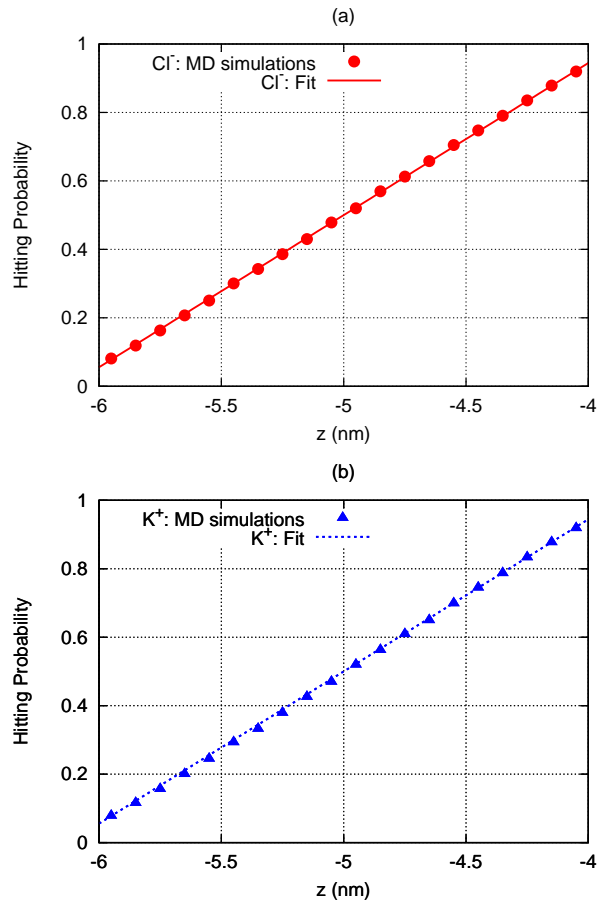}
     \caption{ Hitting probability, $P_+(z_0)$, of reaching the upper boundary of the bulk region (located at $z = -4$nm) for (a) chloride ions (b) potassium ions as a function of their initial position, $z_0$.  The linearity of the curves is an evidence supporting that $F_\text{eff} \simeq 0$.}
      \label{pbulk_HittingProb}
   \end{center}
 \end{figure}

From the calculation of the hitting probability $P_+^{(MD)}(z_0)$ we have verified that, due to the much higher resistance of the membrane+channel, in the bulk there is a negligible drop of electrostatic potential. The effective force $F_\text{eff}$ obtained from the comparison of $P_+^{(MD)}(z_0)$ with Eq.\,(\ref{hitprob}), which is determined by the value of the electric field, is not distinguishable from $0$ (see Fig. (\ref{pbulk_HittingProb})).

\begin{figure}[ht!]
   \begin{center}
        \includegraphics[angle=0,width=8.5cm]{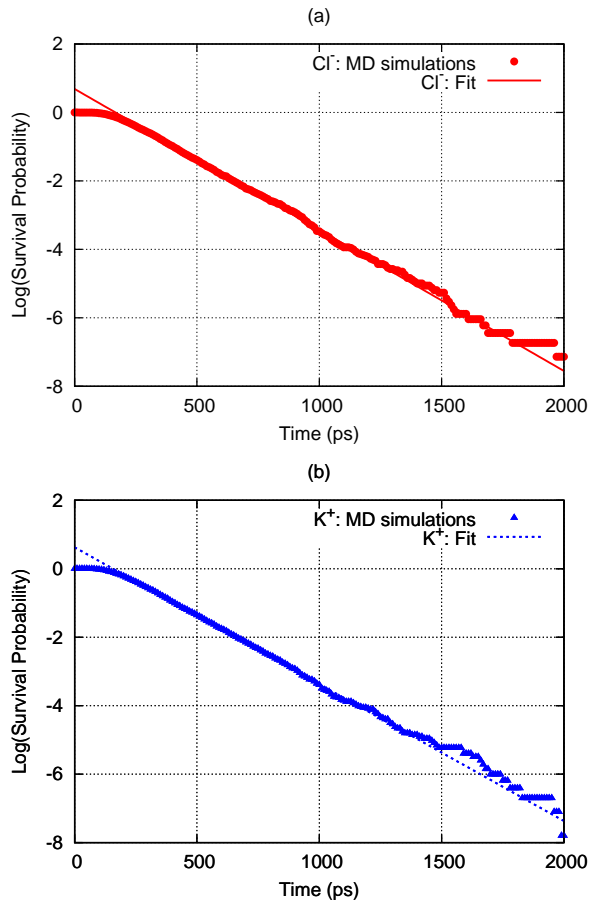}
      \caption{ Survival probability in bulk of (a) chloride ions (b) potassium ions.}
      \label{pbulk_SurvP}
   \end{center}
 \end{figure}

The analysis of the survival probability provides the diffusion coefficients for the Cl$^-$ and K$^+$ ions. From the slope of the fitting straight line represented in Figs.\,(\ref{pbulk_SurvP}) we obtain $D_{Cl} = 2.11$nm$^2/$ns and $D_{K} = 2.05$nm$^2/$ns with the help of Eq.\,(\ref{SurvP}). 

These values compare very well with the values obtained from the study of the mean exit time. By fitting the results for the mean exit time computed from the MD simulations, $T^{(MD)}(z_0)$, with its corresponding analytical expression, Eq.\,(\ref{MET}), we obtain $D_{Cl} = 2.08$nm$^2/$ns and $D_{K} = 2.07$nm$^2/$ns. Such agreement of the parameters, obtained from two independent methods, shows the consistency and robustness of the FPT procedure employed to extract the values of the diffusion coefficient $D$ and effective force $F_\text{eff}$ that characterize the simple 1D model. 
An extra independent test was done to check the reliability of the results obtained from a first passage time approach. Simulations of a 1M KCl electrolyte in a box with periodic boundary conditions were performed and calculated the diffusion coefficients using Einstein's formula for the MSD \cite{Molsim}. The resulting diffusion coefficients are $D_{Cl} = 2.07$nm$^2/$ns and $D_{K} = 2.08$nm$^2/$ns, which are in very good agreement with both sets of values obtained from a first passage time approach. The results are also in good agreement with the experimental bulk diffusion coefficients for $K^+$ and $Cl^-$ in a 1M KCl concentration at 298K: $D_{Cl}^{(exp)} = 2$ nm$^2/$ns and $D_{K}^{(exp)} = 1.85$ nm$^2/$ns \cite{Mills}.

\begin{figure}[ht!]
   \begin{center}
        \includegraphics[angle=0,width=8.5cm]{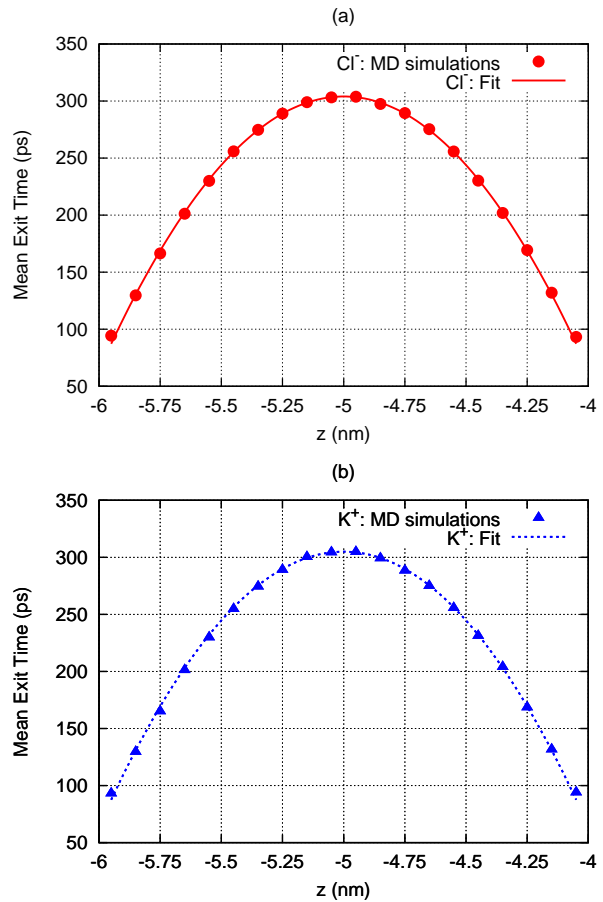}
      \caption{Mean exit time in bulk of (a) chloride ions, (b) potassium ions as a function of their initial position $z_0$.}
      \label{pbulk_MET}
   \end{center}
 \end{figure}

\subsection{Channel}

\begin{figure}[ht!]
   \begin{center}
        \includegraphics[angle=-90,width=8.5cm]{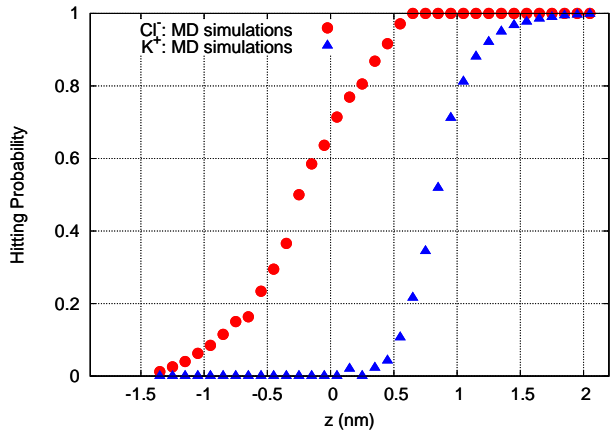}
     \caption{ Hitting probability, $P_+(z_0)$, of ions reaching the upper boundary of the channel (located at $z = 2.1$nm) as a function of the initial position of the particle, $z_0$.}
      \label{pall_HittingProb}
   \end{center}
 \end{figure}

\begin{figure}[ht!]
   \begin{center}
        \includegraphics[angle=-90,width=8.5cm]{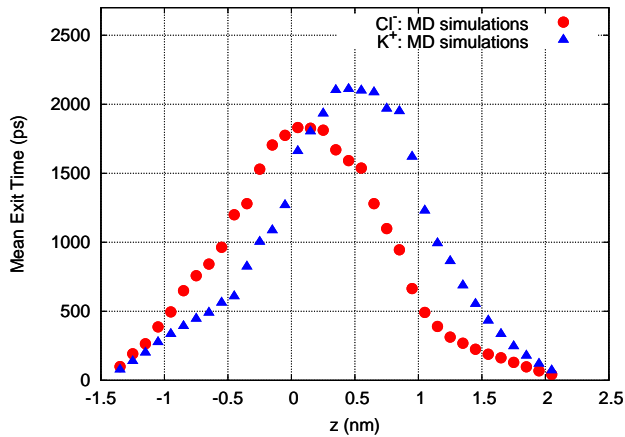}
      \caption{Mean exit time of ions in the channel as a function of their initial position $z_0$.}
      \label{pall_MET}
   \end{center}
 \end{figure}

\begin{figure}[ht!]
   \begin{center}
      \includegraphics[width=8.5cm]{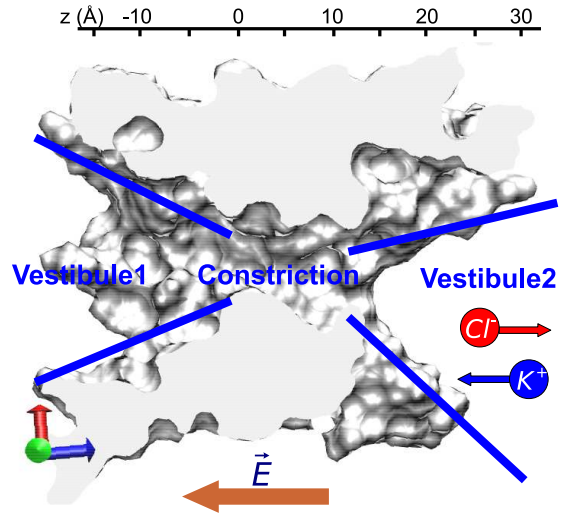}
      \caption{(Color online) Different regions of the system: Vestibule1,  $-1.4$ nm$ < z < -0.3$ nm. Constriction, $-0.3$ nm$ < z < 1.1$ nm. Vestibule2, $1.1$ nm$ < z < 2.1$ nm. Figure produced using VMD \cite{VMD}.}
      \label{regions}
   \end{center}
 \end{figure}

In the following, we present the results for the First Passage quantities corresponding to the transport of ions across the whole channel. The channel is defined as the central part of the OmpF trimer in which the monomers have a rigid beta barrel structure, leaving out the more flexible loops and turns located at the edges of the protein. In our simulation, the channel approximately extends from $z = -1.4$ nm to $z = 2.1$ nm, see Fig. (\ref{regions}). The results for the FPT quantities, Figs.\,(\ref{pall_HittingProb}) and (\ref{pall_MET}), are obtained from the analysis of the trajectories of the ions of the all-atom MD simulation of the system, which contains the exact geometry and considers all interactions within the CHARMM force field. They provide valuable information on the ion transport inside the channel. From Fig.\,(\ref{pall_HittingProb}) we can infer the existence of an energy barrier located in the vicinity of $z \sim 0.5$ nm that prevents the passage of both ions, since the probability of crossing $z \sim 0.5$ nm is very small. The results on the MET represented in Fig.\,(\ref{pall_MET}) provide the characteristic residence times for both ions as a function of their position within the channel. From those values we can obtain the average time of residence in the channel for both ions, $\overline{T}_{Cl} =822$ ps for chloride and $\overline{T}_{K} = 935$ ps for potassium ions. The different position of the maximum of the MET for both ions, located at $z<z_{L/2}$ for chloride ions and at $z>z_{L/2}$ for potassium ions (with $z_{L/2}$ being the middle of the channel), reflects the overall action of the external electric field (directed towards negative $z$-direction), which drags cations along the field and anions in the opposite direction. 

Another interesting outcome of the analysis is that the results for the hitting probabilities and mean exit times shown in Figs. (\ref{pall_HittingProb}) and (\ref{pall_MET}), contrary to the bulk case, cannot be satisfactorily fitted by the analytical expressions Eqs.\,(\ref{hitprob}) and (\ref{MET}). Our results show that it is not possible to assign constant diffusion coefficients to the ions inside the channel. Therefore, this result cast some doubt on the adequacy of one-dimensional models based on the Nernst-Planck equation with constant diffusion coefficients to describe the ion transport through the whole OmpF porin.

In order to reach a better understanding of the ionic transport through the OmpF porin, we divided the system into different regions and, through the comparison of the FPT quantities, $P_{\pm}^{(MD)}(z_0), T^{(MD)}(z_0), S_p^{(MD)}(t)$, with the corresponding analytical expressions, determined the parameters that characterize the ion dynamics (i.e., the effective diffusion coefficient $D$ and the effective force $F_\text{eff}$) in each region. Based on its structure, we split the channel into three different regions: {\it Constriction}, defined as the central region between $z = -0.3$ nm and $z = 1.1$ nm; {\it Vestibule1}, defined by  $z = -1.4$ nm and $z = -0.3$ nm; and {\it Vestibule2}, in between  $z = 1.1$ nm and $z = 2.1$ nm (see Fig.\,(\ref{regions})).

\subsection{Channel Constriction}

\begin{figure}[ht!]
   \begin{center}
      \includegraphics[angle=-90,width=8.5cm]{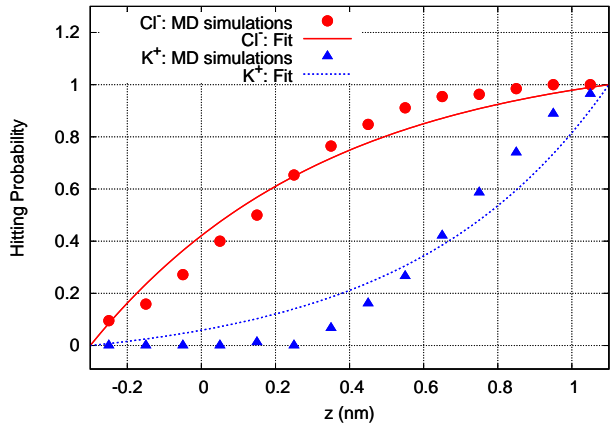}
      \caption{Hitting probability, $P_+(z_0)$, of ions reaching the upper boundary of the constriction (located at $z = 1.1$nm) as a function of the initial position of the particle, $z_0$. The concavity of the curve in the chloride ion case indicates the action of a force along the $z$-direction, $F_\text{eff}^{(Cl)} =  1.6$ k$_B$T$_r$/nm. The convexity of the curve in the potassium ion case indicates the action of a force in the opposite direction, $F_\text{eff}^{(K)} = - 1.9$ k$_B$T$_r$/nm.}
      \label{pconstr_HittingProb}
   \end{center}
 \end{figure}

\begin{figure}[ht!]
   \begin{center}
      \includegraphics[angle=270, width=8.5cm]{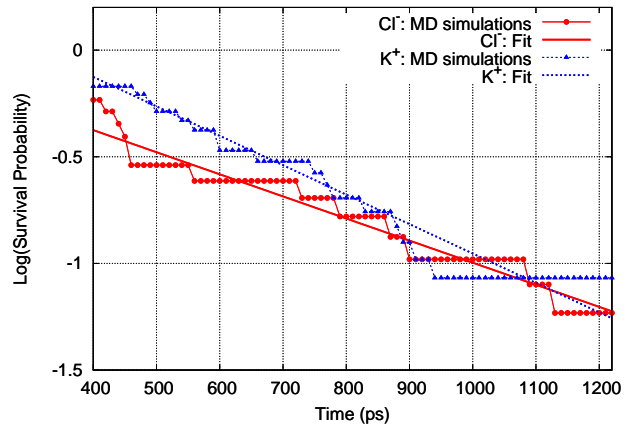}
      \caption{Survival probability of ions in the constriction.}
      \label{pconstr_SurvP}
   \end{center}
 \end{figure}

\begin{figure}[ht!]
   \begin{center}
      \includegraphics[angle = -90, width=8.5cm]{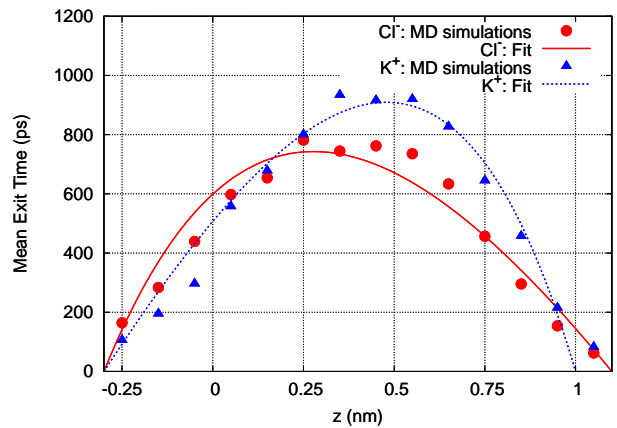}
      \caption{Mean exit time of ions in the constriction as a function of their initial position.}
      \label{pconstr_MET}
   \end{center}
 \end{figure}

We employed the same protocol used in the analysis of the ions in bulk to study the dynamics of $K^+$ and $Cl^-$ in the constriction region. Although the quantitative agreement between the first passage quantities computed from the simulations results and the corresponding analytical expressions is not accurate, the general functional form of the MD results is well captured by the analytical expressions with the appropriate constant parameters (see Figs.\,(\ref{pconstr_HittingProb}), (\ref{pconstr_SurvP}), (\ref{pconstr_MET})). Such an agreement ensures the adequacy of the 1D diffusion-advection, Eqs.\,(\ref{cons}, \ref{NP}), to describe the dynamics of the ions inside the constriction. It must be stressed here the reduced number of ions present in the constriction, which results in poor statistics and might contribute to the observed deviations. 

From the analysis of the hitting probabilities, we obtain the values of the effective force acting on the ions (see Fig.\,(\ref{pconstr_HittingProb})). The effective force is $F_\text{eff}^{(K)} = - 1.9$ k$_B$T$_r$/nm and $F_\text{eff}^{(Cl)} =  1.6$ k$_B$T$_r$/nm for potassium and chloride ions, respectively. As expected, potassium ions are dragged along the direction of the external electric field, whereas chloride ions are dragged in the opposite direction. The slightly different magnitude of both forces is consistent with the observation that anions and cations follow different pathways in the constriction, as has been reported in previous MD studies \cite{Roux2002} and in experiment \cite{Roux2010}. Assuming that the section of the channel along the $z$-axis is uniform in the constriction (see Fig. (\ref{regions})), it follows (see below) that the effective force acting on the ions is determined solely by the local electric field acting on them. From the values of the effective forces we can estimate the magnitude of the local electric field for potassium ions, $ |F_\text{eff}^{(K)}/e| = 0.049$ V/nm, and for chloride ions, $|F_\text{eff}^{(Cl)}/e| = 0.041$ V/nm. 

The fit of the time dependence of the survival probabilities obtained from the MD simulations $S_p^{(MD)}(t)$  with Eq.\,(\ref{SurvP}) provides the diffusion coefficients $D_{Cl} =0.18  \pm 0.1$ nm$^2$/ns, $D_K = 0.21 \pm 0.1$ nm$^2$/ns. As seen in Fig. (\ref{pconstr_SurvP}), the statistics in this case is poor, so the results for the diffusion coefficients are associated with large errors. 

These values compare well with the values obtained from the analysis of the mean exit time: $D_{Cl} =0.26$ nm$^2$/ns, $D_K = 0.22$ nm$^2$/ns. Also, as seen in Fig.\,(\ref{pconstr_MET}), the profile of the mean exit time is well reproduced by Eq.\,(\ref{MET}) with such parameters.  Note that the diffusion coefficients of both ions in the constriction are roughly an order of magnitude smaller than the diffusion coefficients in the bulk. 

\subsection{Channel Vestibules}

A similar procedure was employed to analyze the dynamics of the ions at the vestibules of the constriction, defined as the regions between $z = -0.3$nm and $z = -1.4$nm (Vestibule1) and $z = 1.1$nm and $z = 2.1$nm (Vestibule2)  (see Fig. (\ref{regions})).

\begin{figure}[htp]
   \begin{center}
        \includegraphics[angle = 0, width=8.5cm]{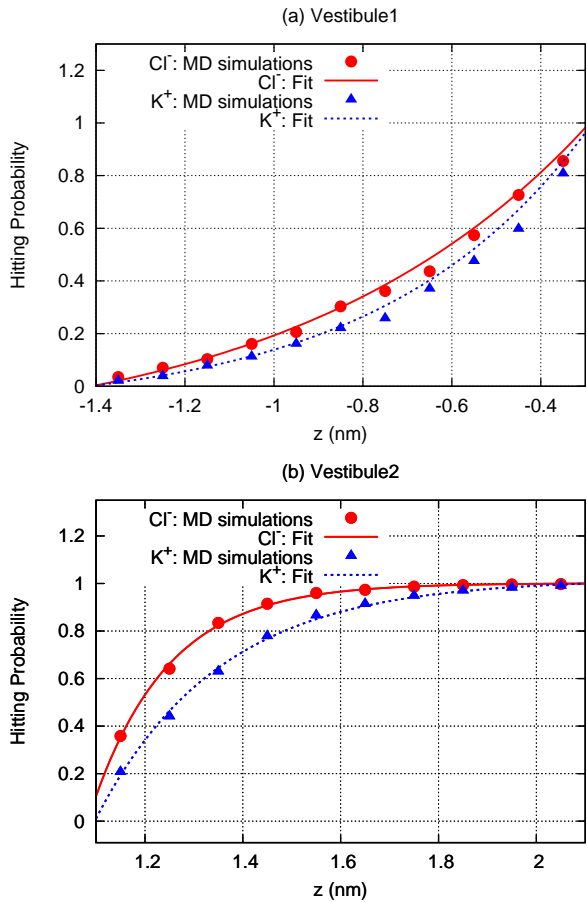}
      \caption{\label{pvest_HittingProb}(a) Hitting probability, $P_+(z_0)$, of ions reaching the upper boundary of Vestibule1 (located at $z = -0.3$nm) as a function of the initial position of the particle, $z_0$. For chloride ions, $F_\text{eff}^{(Cl)} =-1.4$k$_B$T$_r$/nm is used. For potassium ions, $F_\text{eff}^{(K)} =-2.6$k$_B$T$_r$/nm. (b) Hitting probability, $P_+(z_0)$, of ions reaching the upper boundary of Vestibule2 (located at $z = 2.1$nm) as a function of the initial position of the particle, $z_0$. For chloride ions, $F_\text{eff}^{(Cl)} =5.8$k$_B$T$_r$/nm. For potassium ions, $F_\text{eff}^{(K)} =4.0$k$_B$T$_r$/nm.}
   \end{center}
 \end{figure}

\begin{figure}[ht!]
   \begin{center}
        \includegraphics[angle=0, width=8.5cm]{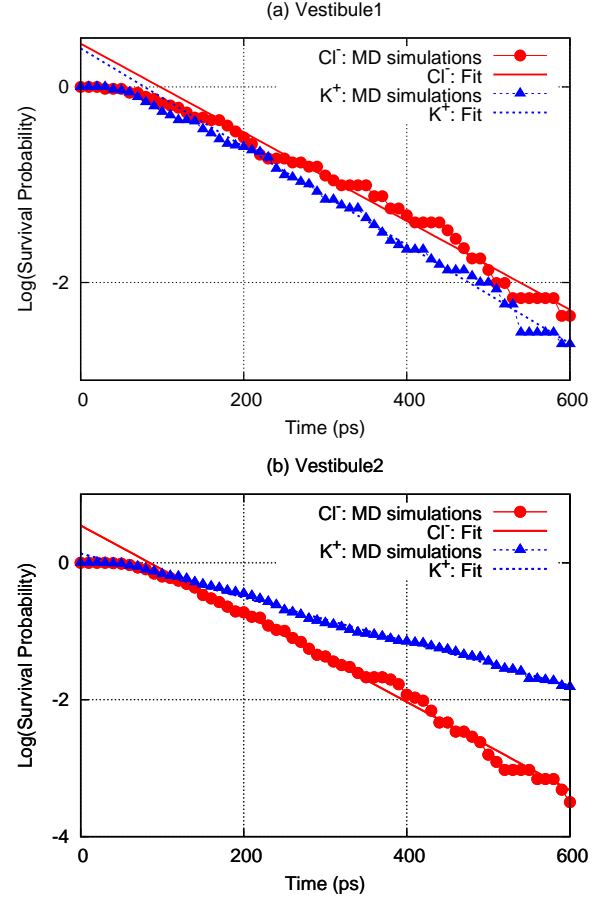}
      \caption{\label{pvest_SurvP} Survival probability of ions in (a) Vestibule1 (b) Vestibule2.}
   \end{center}
 \end{figure}

\begin{figure}[ht!]
   \begin{center}
        \includegraphics[angle=0,width=8.5cm]{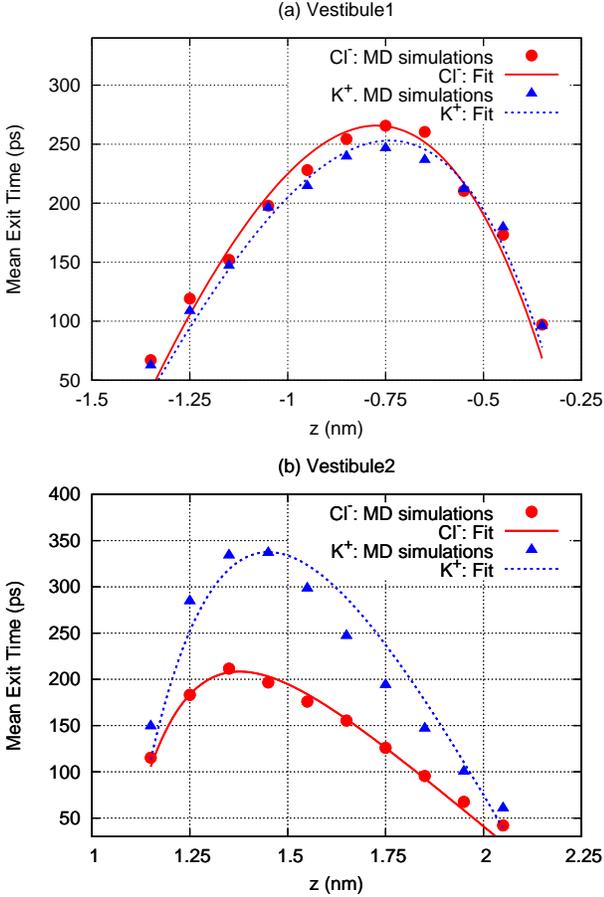}
      \caption{\label{pvest_MET} Mean exit time of ions in (a) Vestibule1 (b) Vestibule2 as a function of their initial position.}
   \end{center}
 \end{figure}

The first general observation is the remarkable good agreement between the first passage quantities computed from the MD simulations results and the corresponding analytical expressions with the appropriate constant parameters (see Figs.\,(\ref{pvest_HittingProb}), (\ref{pvest_SurvP}) and (\ref{pvest_MET})). This indicates that the dynamics of the ions in the vestibules is properly described by the simple 1D model based on Eqs.\,(\ref{cons},\ref{NP}) with the appropriate parameters $D$ and $F_\text{eff}$. 

For the vestibules, the results for the hitting probabilities $P_+^{(MD)}(z_0)$ give the surprising results shown in Figs.\,(\ref{pvest_HittingProb}). In the two regions, the effective foreces acting on both anions and cations have the same sign, being negative in Vestibule1 ($F_\text{eff}^{(Cl)} =-1.4$k$_B$T$_r$/nm for chloride ions, and $F_\text{eff}^{(K)} =-2.6$k$_B$T$_r$/nm for potassium ions) and positive in Vestibule2 ($F_\text{eff}^{(Cl)} = 5.8$k$_B$T$_r$/nm for chloride ions, and $F_\text{eff}^{(K)} = 4.0$k$_B$T$_r$/nm for potassium ions). This indicates that the dominant contribution to the effective force dragging the ions is not charge dependent and its direction must be along the $z$-axis (and of larger magnitude) in Vestibule2 and in the opposite direction in  Vestibule1. Also, the magnitudes of the effective forces acting on each ion are different. The effective force acting on potassium ions is greater than the force on chloride ions in Vestibule1, but it is smaller in Vestibule2. \\

Such results correspond well with the action of charge independent {\it entropic forces}, which arise when the dynamics of particles through channels is described with one dimensional models. In treating channels of varying cross section, the analysis of the dynamics of ions can be reduced, under appropriate plausible hypothesis, from a three-dimensional problem with a complex boundary to a one-dimensional problem of diffusion along the channel axis \cite{Jacobs, Zwanzig, Reguera, Kalinay, Berezhkovskii}. Assuming that the distribution of particles in the planes perpendicular to the axis of the channel is close to its equilibrium distribution, one obtains the modified Ficks-Jacobs equation. Following Reguera and Rubi's formulation \cite{Reguera}, in this approximation the flux of particles, $J_{FJ}$, can be expressed as: 
\begin{equation}\label{JFJ}
 J_{FJ}(z,t) = - D\left[\frac{\partial c(z,t)}{\partial z} + \frac{1}{k_BT_r}\frac{\partial A(z)}{\partial z} c(z,t)  \right]\,.
\end{equation}
Here $c(z,t)$ is the 1D particle concentration (with dimensions of particles per unit length), $A(z) \equiv U(z) - TS(z)$ is a free energy given by the electrostatic energy $U(z) = -q E z$ and the entropy $S(z) = k_B \ln{h(z)}$, where $h(z)$ is the dimensionless cross section of the channel. This entropic term accounts for the variation of the space accessible for the diffusion of the particles, dragging them towards positions $z$ of maximum entropy. Note also that Eq.(\ref{JFJ}) is a particular case of Eq.(\ref{cons},\ref{NP}) with a deterministic force arising from free energy barriers, $F=-\partial A(z)/\partial z$.

From Eq.\,(\ref{JFJ}) we can infer that the effective forces obtained in the vestibules from the analysis of first passage quantities can be decomposed into two additive contributions: 
\begin{equation}\label{Feff}
 F_\text{eff} = F_{el} + F_{entr}\,.
\end{equation}
Here, $F_{el}$ is the component of the force arising from electrostatic interactions, whose sign depends on the charge of the ion under consideration. $F_{entr}$ accounts for the effective entropic force, which depends solely on the shape of the channel and which points towards the direction that maximizes the available space. 

\begin{figure}[ht!]
   \begin{center}
        \includegraphics[angle=0, width=8.5cm]{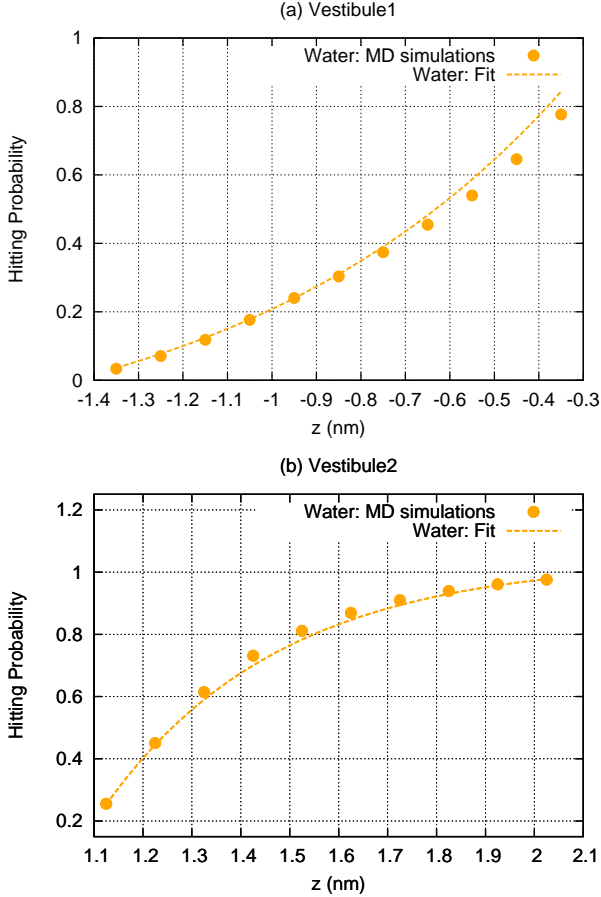}
      \caption{\label{pvestaigua_HittingProb}  Hitting probability, , $P_+(z_0)$, of reaching the upper boundary for water particles in (a) Vestibule1 (b) Vestibule2.}
   \end{center}
 \end{figure}

With the help of Eqs.\,(\ref{Feff}), we can obtain the two contributions, $F_{el}$ and $F_{entr}$, from the values of the total effective force computed from the first passage analysis,
\begin{eqnarray}
 F_{el}^{(K/Cl)} & = & \pm \frac{1}{2}\left(F_\text{eff}^{(K)} - F_\text{eff}^{(Cl)} \right) \nonumber \\
 F_{entr} & = &  \frac{1}{2}\left(F_\text{eff}^{(K)} + F_\text{eff}^{(Cl)} \right)\,.
\end{eqnarray}
In Vestibule1, the resultant effective electrostatic force is $F_{el}^{(K/Cl)} \simeq \mp 0.6$ k$_B$T$_r$/nm, and the entropic part is $F_{entr} = -2$k$_B$T$_r$/nm. In Vestibule2, the resultant electrostatic component is $F_{el}^{(K/Cl)} \simeq \mp 0.9$ k$_B$T$_r$/nm, and the entropic part is $F_{entr} = 4.9$ k$_B$T$_r$/nm.

To confirm the existence of a charge independent entropic force arising from the variation of the available space along the axis of the channel, we have also analyzed the first passage properties of water molecules at both vestibules (see Figs.\,(\ref{pvestaigua_HittingProb})). In this case, the only contribution to the effective force is entropic. For Vestibule1 we find $F_{entr} = -1.5$ k$_B$T$_r$/nm, and for Vestibule2 we find $F_{entr} = 4$ k$_B$T$_r$/nm.

At this point we can interpret qualitatively the results obtained from the analysis of the hitting probabilities at the two vestibules. In Vestibule1, the effective section of the channel decreases along the $z$-axis, which induces the appearance of an {\it entropic force} in the same direction as the external electric field (antiparallel to the $z$-axis). Thus, for cations both contributions to the effective force have the same direction whereas for anions they have opposite directions. As the entropic part is dominant, the effective force on both ions is directed towards decreasing $z$ values, although its magnitude is larger on potassium ions than on chloride ions. In Vestibule2, on the contrary, the effective section of the channel increases along the $z$-axis, so the {\it entropic force} is directed towards increasing $z$ values, and the two contributions of the effective force have the same direction for anions and opposite directions for cations. Hence, in this case the total effective force is in the direction of the $z$-axis and its magnitude is larger for chloride ions than for potassium ions.

The entropic force can be independently estimated using Eqs.\,(\ref{JFJ}) from the dependence of the effective radius of the channel on the $z$-coordinate, as computed from the MD simulations. We obtain an average $F_{entr} = -1.1$ k$_B$T$_r$/nm at Vestibule1 and an average $F_{entr} = 3.1$ k$_B$T$_r$/nm at Vestibule2. The discrepancies in the precise values obtained for the entropic forces were expected, since the expressions used are only valid for smoother changes of the section of the channel along the $z$-coordinate.

Once the effective forces have been determined we can obtain the values of the diffusion coefficients. The logarithmic fit of the time dependence of the survival probabilities, $S_p^{(MD)}(t)$, provides the diffusion coefficients $D_{Cl} =0.65$ nm$^2$/ns and $D_K = 0.66$ nm$^2$/ns in Vestibule1, and $D_{Cl} =0.49$ nm$^2$/ns, $D_K = 0.33$ nm$^2$/ns in Vestibule2 (see Figs.\,(\ref{pvest_SurvP})). Independently, from the analysis of the mean exit times, we obtain $D_{Cl} = 0.57$ nm$^2$/ns, $D_K = 0.59$ nm$^2$/ns in Vestibule1 and $D_{Cl} =0.44$ nm$^2$/ns, $D_K = 0.31$ nm$^2$/ns in Vestibule2 (see Figs.\,(\ref{pvest_MET})). Again, such an agreement of the parameters, obtained from two independent methods, shows the consistency and robustness of the FPT procedure employed to extract the values of the diffusion coefficient $D$.

\section{Conclusions}
\label{sec:Conclusions}
In this paper we have studied the diffusive dynamics of chloride and potassium ions in the interior of the OmpF porin under the influence of an external electric field directed along its axis. From the results of extensive all-atom molecular dynamics simulations of the system we computed several first passage time quantities to characterize the dynamics of the ions in the interior of the channel. Such quantities, calculated considering all the complexity  of the system, provide valuable information on the residence times and presence of free energy barriers for ions in the interior of the OmpF porin. These FPT quantities were also used to test the validity of simplified 1D descriptions of the dynamics of the ions in the interior of the channel based on the use of constant diffusion coefficients for the ions.

Overall, our results show a slow and complex dynamics for the ions inside the channel. The sluggish ionic dynamics inside the channel is characterized by large residence times ($\sim$ 0.8 ns for Cl$^-$ and $\sim 0.9$ ns for K$^+$) and very large mean exit times with peaks around $\sim 2$ ns, see Figure \ref{pall_MET}, which indicate a  strong interaction of the channel with the ions. The complexity of the dynamics is demonstrated by the fact that it is not possible to characterize chloride and potassium ions in the interior of the channel with a single diffusion coefficient. However, it is possible to find well-defined diffusion coefficients for some regions of the protein channel.

Based on its X-ray structure, we have divided the channel into three different regions: constriction, Vestibule 1 and Vestibule 2. Employing the same FPT analysis we demonstrate that each of such regions is describable by a simple model characterized by a constant diffusion coefficient $D$ and a constant deterministic force $F_\text{eff}$ acting on the ions. The diffusion coefficients $D$ which characterize the dynamics of chloride and potassium ions in the different regions, obtained by two independent methods, are given in Table \ref{Table:Results}. Let us remark that these diffusion coefficients vary by almost an order of magnitude from the bulk to the interior of the constriction. The physical origin of the effective force $F_\text{eff}$ is also different in the different regions of the channel. The effective forces on the ions at the constriction, as inferred from the FPT results of the MD simulations, correspond to the presence of an intense local electric field directed along the applied electric field which drags potassium ions along and chloride ions in the opposite direction. In the vestibules, the effective forces observed on the ions are consistent with the expected effect of entropic forces which arise from the variation of the available space along the axis of the channel (see Eq.(\ref{JFJ})). 

\begin{table}
\caption{Diffusion coefficients in the different regions. Two values are provided, one inferred from the analysis of the survival probability and the other from the study of the mean exit time.}
\label{Table:Results}
\centering
\begin{tabular}{cc|c|c|}
\cline{3-4}
& & \multicolumn{2}{|c|}{Diffusion coefficient D (nm$^2$/ns)} \\ \cline{1-4}
\multicolumn{1}{|c|}{Region} & \multicolumn{1}{|c|}{Ion} & Survival Prob. & Mean Exit Time  \\ \cline{1-4}
\multicolumn{1}{|c|}{\multirow{2}{*}{Bulk}} &
\multicolumn{1}{|c|}{K$^+$}& 2.05  & 2.07       \\ 
\multicolumn{1}{|c|}{}                        &
\multicolumn{1}{|c|}{Cl$^-$}& 2.11 & 2.08   \\ \cline{1-4}
\multicolumn{1}{|c|}{\multirow{2}{*}{Constriction}} &
\multicolumn{1}{|c|}{K$^{+}$}& 0.21 & 0.22   \\ 
\multicolumn{1}{|c|}{}                        &
\multicolumn{1}{|c|}{Cl$^-$}& 0.18 & 0.26 \\ \cline{1-4}
\multicolumn{1}{|c|}{\multirow{2}{*}{Vestibule1}} &
\multicolumn{1}{|c|}{K$^{+}$}& 0.66  & 0.59    \\ 
\multicolumn{1}{|c|}{}                        &
\multicolumn{1}{|c|}{Cl$^-$}& 0.65 & 0.57 \\  \cline{1-4}
\multicolumn{1}{|c|}{\multirow{2}{*}{Vestibule2}} &
\multicolumn{1}{|c|}{K$^{+}$}& 0.33 & 0.31  \\ 
\multicolumn{1}{|c|}{}                        &
\multicolumn{1}{|c|}{Cl$^-$}& 0.49 & 0.44  \\ \cline{1-4}

\end{tabular}
\end{table}

We believe that our results have important implications for the modeling of ionic channels. In this field, there is a great interest in having simple, 1D models, with the capability of predicting the essential features of these extremely complex systems. Such models (like, e.g., the PNP approach) are based on the classical 1D transport equations \cite{Hille} (see Eq.(\ref{cons},\ref{NP})) and heavily rely on the existence of a well-defined diffusion coefficient. Our work strongly suggests that in general a direct use of a constant diffusion coefficient for the whole channel is not justified.  However, it also shows the plausibility of a multi scaling approach, in which these models would play an important role. The idea is to take into account the complexity of the structure of the protein by obtaining diffusion coefficients from detailed MD simulations in regions of different geometrical properties. The coefficients thus obtained can be used in the classical 1D descriptions of ionic channels, such as the PNP approach.  Also, we recall that our work further demonstrates that in such 1D descriptions, entropic forces such as those described in Refs.\cite{Jacobs, Zwanzig, Reguera, Kalinay, Berezhkovskii} must be taken into account. Work in this direction of a multi-scaling approach is now under way.

\section{Acknowledgments}
This work is supported by the Spanish Government (grants FIS2009-13370-C02-02, FIS2007-60205 and CONSOLIDER-NANOSELECT-CSD2007-00041), Generalitat de Catalunya (2009SGR164) and Fundaci\'o Caixa Castell\'o-Bancaixa (P1-1A2009-13). C.C. is supported by the JAE doc program of the Spanish National Research Council (CSIC). JF acknowledges the award of an Intramural PIE grant from the Spanish National Research Council (CSIC). The Supercomputing resources employed in this work were provided by the CESGA Supercomputing Center, Spain.


\begin{thebibliography}{99}
\bibitem{Hille}{B. Hille}, {\it Ion Channels of Excitable Membranes} (Sinaure, Sunderland, 2001).
\bibitem{Maglia}{G. Maglia, A. J. Heron, W. L. Hwang, M. A. Holden, E. Mikhailova, Q. Li, S. Cheley and H. Bayley}, {\it Nature Nanotechnology} {\bf 4}, 437 (2009).
\bibitem{Cowan} {Cowan, S. W., R. M. Garavito, J. N. Jansonius, J. A. Jenkins, R. Karlsson, N. König, E. F. Pai, R. A. Pauptit, P. J. Rizkallah, J. P. Rosenbusch, G. Rummel and T. Schirmer}, {\it Structure} {\bf 3}, 1041 (1995).  
\bibitem{Jacobs}{M. H. Jacobs}, {\it Diffusion Processes} (Springer, New York, 1967).
\bibitem{Zwanzig}{R. Zwanzig}, {\it J. Phys, Chem.} {\bf 96}, 3926 (1992).
\bibitem{Reguera}{D. Reguera and J. M. Rub\'i}, {\it Phys. Rev. E} {\bf 64}, 061106 (2001). {D. Reguera, G. Schmid, P. S. Burada, J. M. Rub\'i, P. Reimmann, and P. H\"anggi}, {\it Phys. Rev. Lett.} {\bf 96}, 130603 (2006).
\bibitem{Chung}{ S. Chung and B. Corry}, {\it Soft Matter} {\bf 1}, 417 (2005).
\bibitem{Chen1997}{D. Chen, J. Lear, and R. Eisenberg}, {\it Biophys. J.} {\bf 72}, 97 (1997).
\bibitem{Chen1999}{D. Chen, L. Xu, A. Tripathy, G. Meissner, and R. Eisenberg}, {\it Biophys. J.} {\bf 76}, 1346 (1999).
\bibitem{Tang}{J. Tang, D. Chen, N. Saint, J. Rosenbusch, and R. Eisenberg}, {\it Biophys. J.} {\bf 72}, A108 (1997).
\bibitem{Roux1}{W. Im and B. Roux}, {\it J. Mol. Biol.} {\bf 322}, 851 (2002).
\bibitem{Lidon} {M. Lid\'on L\'opez, M. Aguilella-Arzo, V. M. Aguilella, A. Alcaraz}, {\it J. Phys. Chem. B} {\bf 113}, 8745 (2009).
\bibitem{Aksimentiev} {Aksimentiev, A. and K. Schulten}, {\it Biophys. J.} {\bf 88}, 3745 (2005).
\bibitem{Ulrich2009} { Pezeshki, S., C. Chimerel, A. N. Bessonov, M. Winterhalter and U. Kleinekathofer}, {\it Biophys. J.} {\bf 97}, 1898 (2009).
\bibitem{Faraudo}{J. Faraudo, C. Calero and M. Aguilella-Arzo}, {\it Biophys. J.} {\bf 99}, 2107 (2010). 
\bibitem{Roux2001}{ W. Im and B. Roux}. {\it J. Chem. Phys.} {\bf 115}, 4850 (2001).
\bibitem{Roux2002b}{ W. Im and B. Roux}. {\it J. Mol. Biol.} {\bf 322}, 851 (2002).
\bibitem{Schirmer}{T. Schirmer and P. S. Phale}, {\it J. Mol. Biol.} {\bf 294}, 1159 (1999).
\bibitem{Phale}{ P.S. Phale, A. Philippsen, C. Widmer, V. P. Phale, J. P. Rosenbusch, and T. Schirmer}, {\it  Biochemistry} \textbf{40}, 6319 (2001)
\bibitem{Bolintineanu}{ D. S. Bolintineanu, A. Sayyed-Ahmad, H. T. Davis, Y. N. Kaznessis}, { \it PLoS Comput. Biol.} {\bf 5}, 1 (2009).
\bibitem{Tieleman}{D. P. Tieleman and H. J. C. Berendsen}, {\it Biophys. J} {\bf 74}, 2786 (1998).
\bibitem{Roux2002} {W. Im and B. Roux}, {\it J. Mol. Biol.} {\bf 319}, 1177 (2002).
\bibitem{Allen}{T. W. Allen, S. Kuyucak and S.-H. Chung}, {\it Biophys. J} {\bf 77}, 2502 (1999).
\bibitem{Hijkoop}{V. J. van Hijkoop, A. J. Dammers, K. Malek, and M. O. Coppens}, {\it J. Chem. Phys.} {\bf 127}, 085101 (2007).
\bibitem{Munakata}{T. Munakata and Y. Kaneko}, {\it Phys. Rev. E} {\bf 47}, 4076 (1993).
\bibitem{VanKampen}{N. G. van Kampen}, {\it Stochastic precesses in physics and chemistry}, (North-Holland publishing company, 1981).
\bibitem{Redner}{S. Redner}, {\it A Guide to First-Passage Processes}, (Cambridge University Press, 2001).
\bibitem{PDB} {Cowan, S.W.} The Refined Structure of Ompf Porin from E.Coli at 2.4 Angstroms Resolution. \emph{Protein Data Bank} freely available at http://www.pdb.org code 2OMF (DOI:10.2210/pdb2omf/pdb).
\bibitem{Rostovtseva2002} {T.K. Rostovtseva, E.M. Nestorovich and S. M. Bezukov}, {\it Biophys. J.} \textbf{82}, 160 (2002).
\bibitem{Danelon2006} {C. Danelon, E. M. Nestorovich, M. Winterhalter, M. Ceccarelli, S. M. Bezrukov}{\it Biophys. J.} \textbf{90}, 1617 (2006).
\bibitem{Marcel}{M. Aguilella-Arzo, C. Calero and J. Faraudo}, {\it SoftMatter}, DOI: 10.1039/C0SM00904K (2010).
\bibitem{VMD} {Humphrey, W., A. Dalke, A. and K. Schulten}, {\it Molec. Graphics} {\bf 14}, 33 (1996).
\bibitem{Varma} {Varma, S., S. Chiu and E. Jakobsson}, {\it Biophys. J.} {\bf 90}, 112 (2006).
\bibitem{NAMD} {Phillips, J. E., R. Braun, W. Wang, J. Gumbart, E. Tajkhorshid, E. Villa, C. Chipot, R.D. Skeel, L. Kal\'e, and K. Schulten}, {\it J. Comp. Chem} {\bf 26}, 1781 (2005).
\bibitem{Charmm} MacKerell, A.D., Brooks, B., Brooks, C.L., III, Nilsson, L. ,Roux,  B., Won, Y. and Karplus, M. CHARMM: The Energy Function and Its Parameterization with an Overview of the Program, in The Encyclopedia of Computational Chemistry, 1, 271-277, P. v. R. Schleyer et al., editors, Ed. John Wiley \& Sons, Chichester, 1998. 
\bibitem{Marshall}{T. W. Marshall and E. J. Watson}, {\it J. Phys. A} {\bf 20}, 1345 (1987).
\bibitem{Molsim} {C. Calero, J. Faraudo, and M. Aguilella-Arzo}, {\it Molecular Simulation} (accepted). Preprint available at http://arxiv.org/abs/1005.2857 (2010)
\bibitem{Mills}{R. Mills and V. Lobo}, {\it Self-diffusion in Electrolyte Solutions}, Elsevier, Amsterdam (1989).
\bibitem{Roux2010}{Dhakshnamoorthy, B., S. Raychaudhury, L. Blachowicz and B. Roux}, {\it J. Mol. Biol.} {\bf 396}, 293 (2010).
\bibitem{Kalinay}{P. Kalinay, J. K. Percus}, {\it J. Chem. Phys} {\bf 122}, 204701 (2005). {P. Kalinay, J. K. Percus}, {\it Phys. Rev. E} {\bf 74}, 041203 (2003).
\bibitem{Berezhkovskii}{A. M. Berezhkovskii, M. A. Pustovoit, and S. M. Bezrukov}, {\it Phys. Rev. E} {\bf 80}, 020904 (2009).













\end{thebibliography}
\end{document}